\newcounter{myctr}

\documentclass{article}
\usepackage{url,graphicx}
\usepackage{amssymb,amsfonts,amsmath}
\usepackage{caption}
\usepackage{subcaption}
\begin{document}

\makeatletter
\def\@biblabel#1{[#1]}
\makeatother

\title{Unstable Dynamics of Adaptation in Unknown Environment due to Novelty Seeking}
\author{Arkady Zgonnikov\thanks{arkady.zgonnikov@gmail.com}, Ihor Lubashevsky\thanks{i-lubash@u-aizu.ac.jp} \\ 
\small{University of Aizu, Tsuruga, Ikki-machi}\\
\small{Aizu-Wakamatsu, Fukushima, 965-8580 Japan}}

\date{}

\maketitle

\begin{abstract}
Learning and adaptation play great role in emergent socio-economic phenomena. Complex dynamics has been previously found in the systems of multiple learning agents interacting via a simple game. Meanwhile, the single agent adaptation is considered trivially stable. We advocate the idea that adopting a more complex model of the individual behavior may result in a more diverse spectrum of macro-level behaviors. We develop an adaptation model based on the reinforcement learning framework extended by an additional processing channel. We scrutiny the dynamics of the single agent adapting to the unknown environment; the agent is biased by novelty seeking, the intrinsic inclination for exploration. We demonstrate that the behavior of the novelty-seeking agent may be inherently unstable. One of the surprising results is that under certain conditions the increase of the novelty-seeking level may cause the agent to switch from the non-rational to the strictly rational behavior. Our results give evidence to the 
hypothesis that the intrinsic motives of 
agents should be paid no less attention than the extrinsic ones in the models of complex socio-economic systems.
\end{abstract}


\section{Introduction}
A cornerstone of the complexity science is the notion of emergence, which implies that high-level dynamical patterns may arise in a large-scale system as a result of nonlinear interactions between simple entities comprising the system. It is often stressed that complex phenomena emerge in such systems despite the simplicity and even primitiveness of the individual elements, solely due to the interactions between them. For this reason most models of complex systems tend to employ the microscopic behavior models of minimal possible complexity. 

However, there is reason to think that the increase of the model complexity on the level of individual elements may result in richer macro-level behavior of the system as a whole. This in turn may help to better capture the diverse phenomena observed in the real-world complex systems. The concept of emergence due to individual complexity already proved useful, e.g., in studying the self-organization in social insects~\cite{deneubourg1999self,roces2002individual}. In the general field of computational social science this issue is also acknowledged, so there is an increasing demand for more advanced and authentic microscopic models of individual agent behavior; ``more mental complexity must be modelled to understand what are the specific mental properties allowing social complexity to be managed and simplified''~\cite{conte2012manifesto}. In the present paper we attempt to face this challenge.

\subsection{Learning and adaptation in complex systems}
Learning and adaptation are among the fundamental underlying principles of non-equilibrium social science; ``the non-equilibrium phenomena generated by learning dynamics are a decisive battlefield for computational social science''~\cite{conte2012manifesto}. Particularly, economics exploits the concept of learning in games as the non-equilibrium alternative to the standard Nash equilibrium approach~\cite{arthur1994inductive,erev1998predicting,fudenberg1998theory}. By adopting the perfect rationality hypothesis, the latter imposes ``heroic assumptions about the knowledge and calculating abilities of the players''~\cite{macy2002learning}. The learning approach addresses this issue by focusing on the adaptive behavior of the boundedly rational agents, who repeatedly play a game and gradually learn efficient strategies. In the game-learning setting players often fail to come up with a strategy (either pure or mixed) leading to the Nash equilibrium, so the learning dynamics on its own becomes vital.

A great deal of literature emphasizes richness and complexity of learning dynamics in games~\cite{galla2009intrinsic,galla2011cycles,lubashevsky2010scale,sato2002chaos,sato2003coupled}. Even the simplest systems of two agents learning to play the rock-paper-scissors game may produce quasiperiodic tori, limit cycles, and deterministic chaos~\cite{sato2002chaos,sato2003coupled}. Still, such complex behavior usually emerges as a result of the agent interaction, while the dynamics of the single agent learning is generally presumed to be trivial. Specifically, the reinforcement learning model used in virtually all relevant studies guarantees the stability of the optimal strategy~\cite{sato2005stability}. In the present study we follow the complementary approach, hypothesizing that the individual adaptation may be intrinsically unstable and therefore might be a source of emergence on its own.

Indeed, in the vast majority of available studies on game theoretical learning the agents are still assumed to be rational in a sense that they act selfishly and optimally within the given setting. In fact, their rationality is bounded only in the sense of having less \emph{a priori} information, but their goal --- to maximize the total payoff throughout the whole process --- remains ultimately rational. In the course of learning the agent behavior is driven only by external factors --- the previously observed actions of other players and the revealed payoffs. In the modern models of adaptation the agents basically lack any kind of non-rational motives, they possess no emotions, desires, or personal preferences.

A fundamental feature of human beings, which may hypothetically have a profound impact on the dynamics of adaptation, is the trait of novelty seeking. Humans exhibit novelty-seeking behavior across a diverse range of circumstances, ranging from choosing a vacation destination~\cite{lee1992measuring} to making critical financial decisions~\cite{kuhnen2009genetic}. According to Cloninger's theory, novelty seeking is one of the four basic human personality traits (with three others being reward dependence, harm avoidance and persistence)~\cite{cloninger2004feeling}. In the context of learning novelty seeking propels humans to actively learn and explore~\cite{deci1985intrinsic,ryan2000self}. Appealing to previous studies on human novelty-seeking behavior, we develop a dynamical model of single-agent adaptation as a richer alternative to standard reinforcement learning.

We model the simple situation of a single agent facing an unknown environment, which is represented by a number of rewarded actions. The agent gradually learns the initially unknown rewards by making a repeated choice between these actions (after the action is selected, the corresponding reward is acquired); in general, the rewards may change over time. Canonically, the ultimate goal of the agent is to maximize the total sum of the rewards gained throughout the process. 

In constructing the model we introduce the additional channel of information processing to reinforcement learning. We then analyze in detail how the adaptation dynamics change when the agent behavior is governed not only by the external factors (that is, rewards), but also by the agent intrinsic motives, namely, the inclination to engage in novel activities. What may be the impact of novelty seeking on the dynamics of learning? Can we expect that the non-rational motives change the agent behavior essentially? Do the intrinsic motives deserve as close attention as the extrinsic ones? We attempt to answer these and some other questions in the rest of the paper.

\section{Model}
\subsection{Reinforcement learning}
We construct a model for the adaptation dynamics of a single agent based on the previously elaborated framework of reinforcement learning. An agent is assumed to make repeated choice between finite number of alternatives (or actions, or options) $x_i,\,i=\overline{1,N}$. Each alternative is associated with the corresponding reward $r_i>0$; generally the rewards are non-stationary: $r_i=r_i(t)$. The agent maintains the estimates $q_i$ of each action quality. At every time step $t_k = k\Delta$ ($k\in \mathbb{N}$, $\Delta$ is the time step duration) the estimates $q_i$ are, first, updated with currently received rewards $r_i$, and, second, subjected to the memory loss effect 
\begin{equation}\label{eq:rlm}
   q_i(t_{k+1}) = q_i(t_k) + r_i(t_{k})\delta_{ii_k} - \frac{\Delta}{T_q}q_i(t_k)\,,
\end{equation}
where the index $i_k$ points to the alternative $x_{i}(t_k)$ chosen at the given time step $t_k$. Constant parameter $T_q$ is the agent memory capacity: the events in the past separated from the present by the time considerably exceeding $T_q$ practically do not affect the agent behavior. At each time step only the actually chosen action receives reinforcement (i.e., foregone payoffs are not revealed), so $\delta_{ii_k} = 1$ for $i=i_k$ and $\delta_{ii_k} = 0$ for $i \neq i_k$ (Kronecker delta).

The agent choice is randomized: the Boltzmann ansatz
\begin{equation}
p_i = \frac{e^{\beta q_i}}{\sum\limits_{j}e^{\beta q_j}},
\label{eq:basic_p}
\end{equation}
with the parameter $\beta$ relates the probability $p_i$ of choosing the alternative $x_i$ to the current estimate of its quality $q_i$. Small values of $\beta$ lead to the uniformly random choice ($p_i=N^{-1}$), while $\beta \to \infty$ yields the deterministic ``winner-take-all'' choice: the action with the highest $q_i$ is always selected.

The discrete-time reinforcement learning is traditionally the most popular framework for describing learning and adaptation processes. However, a continuous-time approximation of model~\eqref{eq:rlm} is more suitable for dynamical analysis. When $\Delta \ll T_q$ and $r_i\beta\ll 1$ for $i=\overline{1,N}$, the learning agent has to repeat the choice many times before getting the stable ``opinion'' about the alternatives, so the following approximation~\cite{lubashevsky2010scale,sato2005stability,sato2003coupled} is valid:
\begin{equation}\label{eq:basic_q}
\dot q_i=r_i \psi(p_i) -\frac{q_i}{T_q}\,.
\end{equation} 
In the case of known foregone payoffs (which is not considered here) $\psi(p_i) \equiv 1$, so the system \eqref{eq:basic_p},\eqref{eq:basic_q} directly results~\cite{borgers1997learning} in the renowned replicator equation for learning dynamics\footnote{
The replicator equation can also be derived for the case of unknown foregone payoffs given that the adaptation dynamics is significantly slower comparing to the agent-environment interactions (for detailed discussion see, e.g.,~\cite{sato2005stability}). 
However, under the latter assumption the foregone payoffs effectively become known, because the agent has enough time to collect the complete information on the current rewards~$r_i$ prior to updating the probabilities~$p_i$. 
}, 
which is also widely employed in evolutionary game theory and population ecology. The foregone payoffs are often concealed in the real-world circumstances~\cite{ho2008individual}, hence, in the present paper we focus solely on this case, so that in Eq.~\eqref{eq:basic_q} 
\begin{equation}\label{eq:psi}
\psi(p_i) = p_i\,.
\end{equation} 

According to Eqs.~\eqref{eq:basic_p},\eqref{eq:basic_q},\eqref{eq:psi}, the better rewarded actions are more valuable for the agent, whereas the alternatives with low rewards $r_i$ are perceived as inferior. However, the estimates $q_i$ of the options with low initial probabilities $p_i$ are updated rarely and, hence, such actions are mostly ignored even if they generate relatively high payoffs. This may cause the adaptation process to stagnate at the initial values of $q_i$ and $p_i$, which in fact has little physical meaning. 

In order to overcome this issue we assume that the obtained rewards $r_i$ are weighted with respect to the frequency of the corresponding action: $r_i \to r_iw(p_i)$, so that the quality estimate of a rarely selected action is effectively remembered until the next choice of this action. The weighting function $w(p)=p^{-1}$ has been used in previous studies (see, e.g.,~\cite{kianercy2012dynamics,leslie2005individual,lubashevsky2010scale}). Still, we believe that the generalized ansatz
\begin{equation}\label{eq:w}
w(p)=p^{-\gamma},\quad \gamma>0 
\end{equation}
is more appropriate, incorporating the conventional one as a special case. The drawback of this approach is that Eqs.~\eqref{eq:basic_q},\eqref{eq:psi},\eqref{eq:w} do not result in the standard replicator dynamics, although we believe this is justified by the extra flexibility of the obtained model. Finally, Eqs.~\eqref{eq:basic_q},\eqref{eq:psi},\eqref{eq:w} yield the governing equations for the agent estimates of quality of options $x_i$
\begin{equation}
\dot q_i=r_ip_i^{1-\gamma}-\frac{q_i}{T_q},\quad i=\overline{1,N}.
\label{eq:q_final}
\end{equation}

\subsection{Extension of the reinforcement learning model}
The reinforcement learning model~\eqref{eq:basic_p},\eqref{eq:q_final} draws on the assumption that, first, the agent forms the estimates $q_i$ of quality of each action using the information on the received rewards, and, second, the choice probabilities are based solely on these estimates. We propose the extension of this model which assumes that, first, the factors other than the tangible rewards may influence the agent choice, and, second, these factors are evaluated via the separate processing mechanism.

\begin{figure}
\begin{center}
\includegraphics[width=0.5\columnwidth]{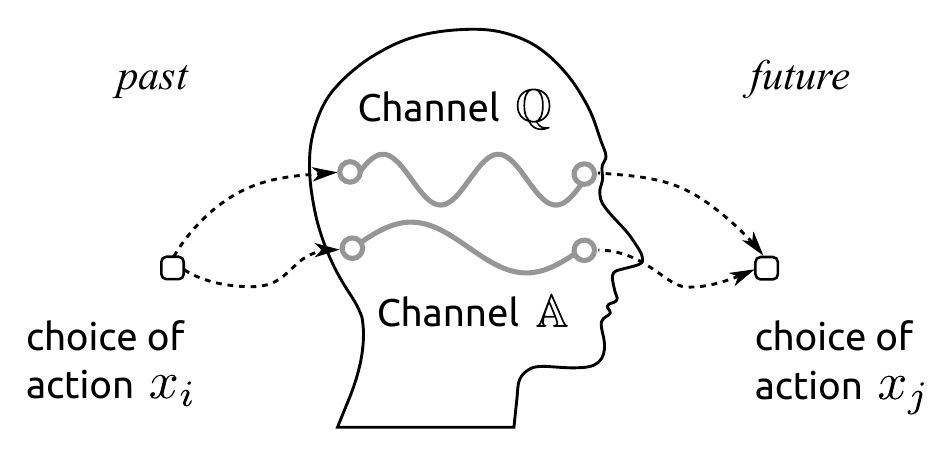}
\end{center}
\caption{Two channels of information processing. Channel $\mathbb{Q}$ which processes the information about the rewards continuously interacts with channel $\mathbb{A}$ which deals with all the information not directly related to the rewards. \label{fig:dual}}
\end{figure}

In evaluating the alternatives during adaptation humans may employ the information not directly related to the objectively available rewards. Such information may include, e.g., cost of taking the action, uncertainty of the agent expectation about the corresponding reward and gain in the knowledge about the environment that can be obtained via the action~\cite{rushworth2008choice}. Accordingly, in our model the agent assesses each option in two different ways. The ``objective'' quality of each action $x_i$ is evaluated through channel $\mathbb{Q}$ and is described by variable $q_i$, whose dynamics essentially depend on the reward function $r_i$~(Eq.~\eqref{eq:q_final}). On the other hand, the agent's estimate of an option based on the information other than the reward is characterized by the separate variable $a_i$, which generally possesses its own dynamics within channel $\mathbb{A}$ (Fig.~\ref{fig:dual}). The pair $\{q_i,a_i\}$ should directly determine the probability $p_i$ of choosing the action $x_i$ 
at each time instant. 

Evaluation of events, conscious and unconscious, is relative, for a discussion see, e.g., \cite{paulus2012emotion}. Within the boundaries of our problem it means that an arbitrary shift individual for the two channels, $q_i \to q_i + C_\mathbb{Q}$ and $a_i \to a_i + C_\mathbb{A}$, where $C_\mathbb{Q}$ and $C_\mathbb{A}$ are some constants, should not affect the probability $p_i$ for any action $x_i$. In physics a dependence $p(q,a)$ meeting this condition is well known, its the Boltzmann or Gibbs distribution 
\begin{equation}\label{eq:BGdistr}
p_i(q_i,a_i) = \frac1Z \exp\big\{\beta_\mathbb{Q} q_i+ \beta_\mathbb{A} a_i\big\}\,.
\end{equation}    
Here $Z$ is the partition function whose list of arguments comprises the quantities $\{q_j,a_j\}$ for all the actions $\{x_j\}$; it has been introduced to normalize the probabilities $\{p_i\}$ to unity. 

The constants $\beta_\mathbb{Q}$, $\beta_\mathbb{A}$ are the characteristics of the set of actions $\{x_j\}$ and should be regarded as system parameters. Their inverse values, $1/\beta_\mathbb{Q}$ and $1/\beta_\mathbb{A}$, actually specify the fuzzy thresholds of the agent perception; when for two actions $x_1$ and $x_2$ the corresponding quantities $\{q_1,a_1\}$ and $\{q_2,a_2\}$ meet the inequalities $|q_1-q_2|\lesssim 1/\beta_\mathbb{Q}$ and $|a_1-a_2|\lesssim 1/\beta_\mathbb{A}$, the agent is not able to distinguish between them in quality, and, consequently, has to regard the two actions as equivalent, so their choice is equiprobable. Measuring the quantities $\{q_i\}$ and $\{a_i\}$ in the units of their thresholds we can set $\beta_\mathbb{Q}=1$ and $\beta_\mathbb{A} =1$, which will be used in the following 
constructions. 

Finally, to complete the model we specify the concrete equations governing the dynamics of the phase variables $a_i$. 

\subsection{Novelty seeking and learning}
Humans often tend to engage in novel activities despite rewards or harm yielded by these activities. Novelty seeking is manifest in many situations and is classified as one of the fundamental human traits~\cite{cloninger1985unified,cloninger2004feeling,cloninger1993psychobiological}. Novelty-seeking behavior may perpetuate the process of learning~\cite{deci1985intrinsic,ryan2000intrinsic,ryan2000self}, and, in addition, may enhance the performance of the learning agent~\cite{oudeyer2007what,oudeyer2007intrinsic}. In the model to be constructed we hypothesize that the agent evaluates the available actions in part based on the novelty of each action, i.e., how often the action has been selected in the recent past.


The novelty of the action $x_i$ can be quantified by the choice probability $p_i$: the actions with low probabilities are considered as novel. On the other hand, it may be interpreted in a sense that the agent becomes bored of the predominant choices, even if they are highly valued in terms of rewards. Subsequently, we propose the following equation governing the dynamics of $a_i$ 
\begin{equation}
\dot a_i=\phi p_i-\frac{a_i}{T_a},
\label{eq:a}
\end{equation}
where $\phi\ge0$ indicates the sensitivity of the agent to the frequencies of the alternatives. To put it another way, parameter $\phi$ quantifies the relative impact of the novelty-seeking trait on the agent choice. Parameter $T_a>0$ determines the agent memory capacity in analogy to $T_q$ in Eq.~\eqref{eq:q_final}.

We wish to draw attention to the fact that, according to Eq.~\eqref{eq:a}, high probability $p_i$ causes $a_i$ to grow, while low $p_i$ leads to decreasing $a_i$. Thus, $a_i$ characterizes in fact not the attraction, but rather the aversion of the agent to the option $x_i$: the higher $a_i$, the stronger the aversion. For this reason in our model the effect of $a_i$ on choice probability $p_i$ is negative (cf. Eq.~\eqref{eq:BGdistr}):
\begin{equation}
p_i(t) = \frac{e^{q_i-a_i}}{\sum\limits_{j}e^{q_j-a_j}}.
\label{eq:p}
\end{equation}
Putting together Eqs.~\eqref{eq:q_final},\eqref{eq:a},\eqref{eq:p}, we finally obtain the model of adaptation of the novelty-seeking agent.
\begin{equation}
\begin{aligned}
\dot q_i&=r_ip_i^{1-\gamma}-\frac{q_i}{T_q},\quad i\in\{1,N\}, \\
\dot a_i&=\phi p_i-\frac{a_i}{T_a},\\
p_i &= \frac{e^{q_i-a_i}}{\sum\limits_{j}e^{q_j-a_j}}.
\end{aligned}
\label{eq:final}
\end{equation}

\section{Dynamics of learning}
In the rest of the paper we analyze the dynamics of the model defined by Eqs.~\eqref{eq:final}. In order to concentrate on the very basic features of the model, we confine our scope to the simplest case of two alternatives $N=2$. Moreover, we assume that the environment changes slowly comparing to the agent adaptation time scale, i.e., the reward function is effectively constant $$r_i(t)\equiv r_i,\quad i=1,2.$$ 

Before examining the properties of the model in case of equal rewards, we denote $r =(r_1+r_2)/2,\,\epsilon =(r_1-r_2)/2$ and transform Eqs.~\eqref{eq:final} in order to simplify the further analysis. Introducing the new variables $q=\frac{q_1-q_2}{2},\quad a=\frac{a_1-a_2}{2}$ and rescaling the time $t\to T_a t,$ we reduce Eqs.~\eqref{eq:final} to the dimensionless system
\begin{equation}
\begin{aligned}
\tau \dot q &=R\frac{\text{sinh}((1-\gamma)(q-a))}{\text{cosh}^{1-\gamma}(q-a)}+\epsilon\frac{\text{cosh}((1-\gamma)(q-a))}{\text{cosh}^{1-\gamma}(q-a)}-q, \\
\dot a &= \Phi\, \text{tanh}(q-a)-a,
\end{aligned}
\label{eq:rescaled}
\end{equation}
where $\tau=\frac{T_q}{T_a}$, $R=r2^{\gamma-1} T_q $ and $\Phi=\frac{\phi}{2} T_a$. Positive values of $q-a$ of order unity correspond to the option $x_1$ preferred over $x_2$ ($p_1\approx1$), and values of $q-a$ close to zero mean that there is practically no difference between the options for the agent ($p_1\approx p_2 \approx 0.5$).

We emphasize that under the adopted assumptions the system~\eqref{eq:rescaled} is equivalent to Eqs.~\eqref{eq:final} and at the same time has less parameters. Hereafter we investigate its properties depending on five parameters: $R$, $\gamma$, $\tau$, $\Phi$ and $\epsilon$. We begin with the detailed analysis of the degenerate case of the equally rewarded actions ($\epsilon=0$), and then examine how the found patterns of behavior change when one of the actions becomes strictly optimal in terms of rewards ($\epsilon > 0$).

\subsection{The case of equal rewards}
Given that $\epsilon=0$, system~\eqref{eq:rescaled} has at least one fixed point $(q=0,\, a=0)$ regardless of the parameter values. Linear stability analysis of this equilibrium reveals that it is always stable given $\gamma\ge1$. In this case the system dynamics is trivial, so in what follows we assume $\gamma \in (0,1)$.

The following conditions are found to be necessary and sufficient for the stability of the system~\eqref{eq:rescaled}, $\epsilon = 0$, at the origin
\begin{equation}
\Phi>\frac{R(1-\gamma)-1}{\tau}-1\quad \text{and} \quad  \Phi>R(1-\gamma)-1.
\label{eq:stability}
\end{equation}

\begin{figure}
\centering
\includegraphics[width=0.7\columnwidth]{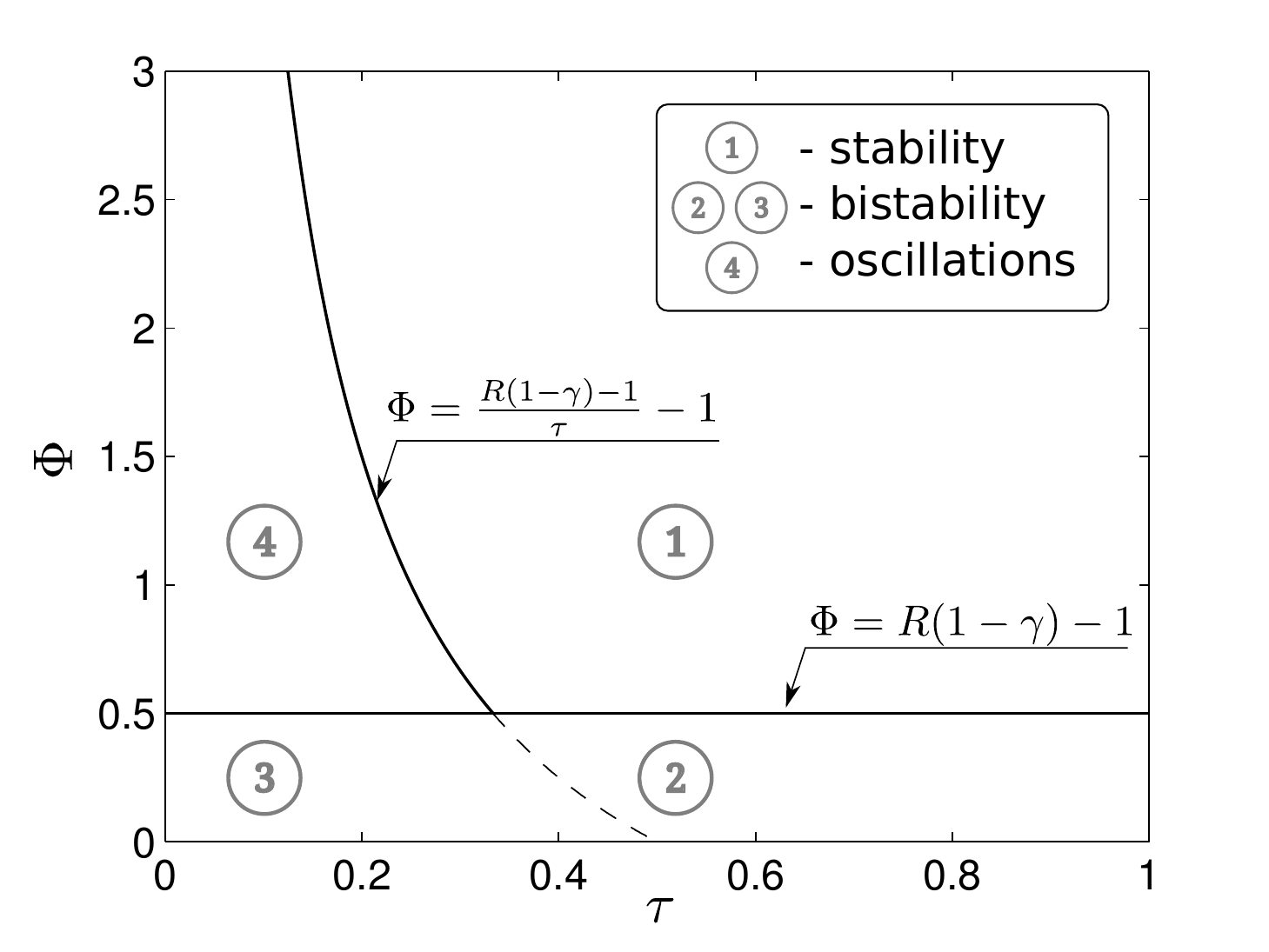}
\caption{Phase diagram of the system~\eqref{eq:rescaled}, $\epsilon = 0$, in the parameter space $(\tau,\Phi)$. The values of two other parameters are fixed: $\gamma=0.5$, $R=3$. Two curves represent the boundaries of system stability (see Eq.~\eqref{eq:stability}). Inside region 1 the system is stable at the origin and has no other attractors. Inside regions 2~and~3 the system has three equilibria: the origin ($q=0$,\,$a=0$) is unstable, whereas ($q=q^*$,\,$a=a^*$) and ($q=-q^*$,\,$a=-a^*$) are both stable. Label 4 denotes the region where the system trajectory forms a limit cycle and has no other stable attractors.\label{fig:stability}}
\end{figure}

Fig.~\ref{fig:stability} represents the curves defined by inequalities~\eqref{eq:stability} for some fixed values of $R$ and $\gamma$. Given that $R>\frac{1}{1-\gamma}$, the curves and two coordinate axes form four regions in the system parameter space. Alternatively, when this assumption is violated, regions 2 and 3 vanish, but inside the remaining regions the patterns of the system dynamics persist. We wish to highlight that under the adopted assumptions ($\gamma \in (0,1)$ and $R>\frac{1}{1-\gamma}$) the basic properties of the system behavior do not depend on $\gamma$ and $R$. Hence we perform the detailed analysis of the system behavior for some fixed values of these parameters (namely, $\gamma=0.5,\,R=3$) without any loss of generality.

The region of the system phase diagram under label 1 (Fig.~\ref{fig:stability}) corresponds to the stability of the fixed point ($q=0,\,a=0$), while in the rest of the parameter space the system is unstable at the origin, so the system dynamics may potentially be non-trivial.

\begin{figure*}
\centering
\includegraphics[width=0.7\columnwidth]{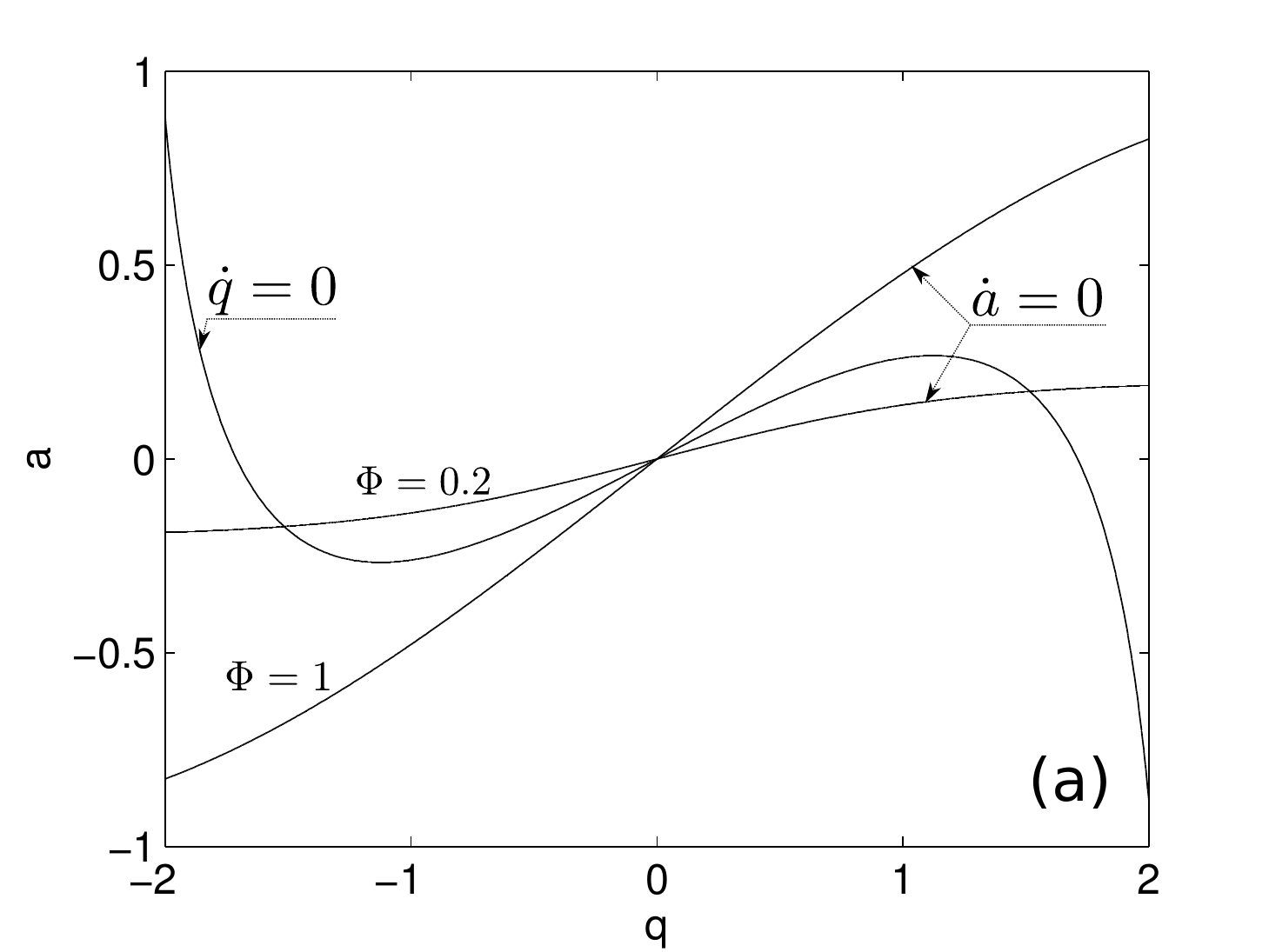}
\includegraphics[width=0.7\columnwidth]{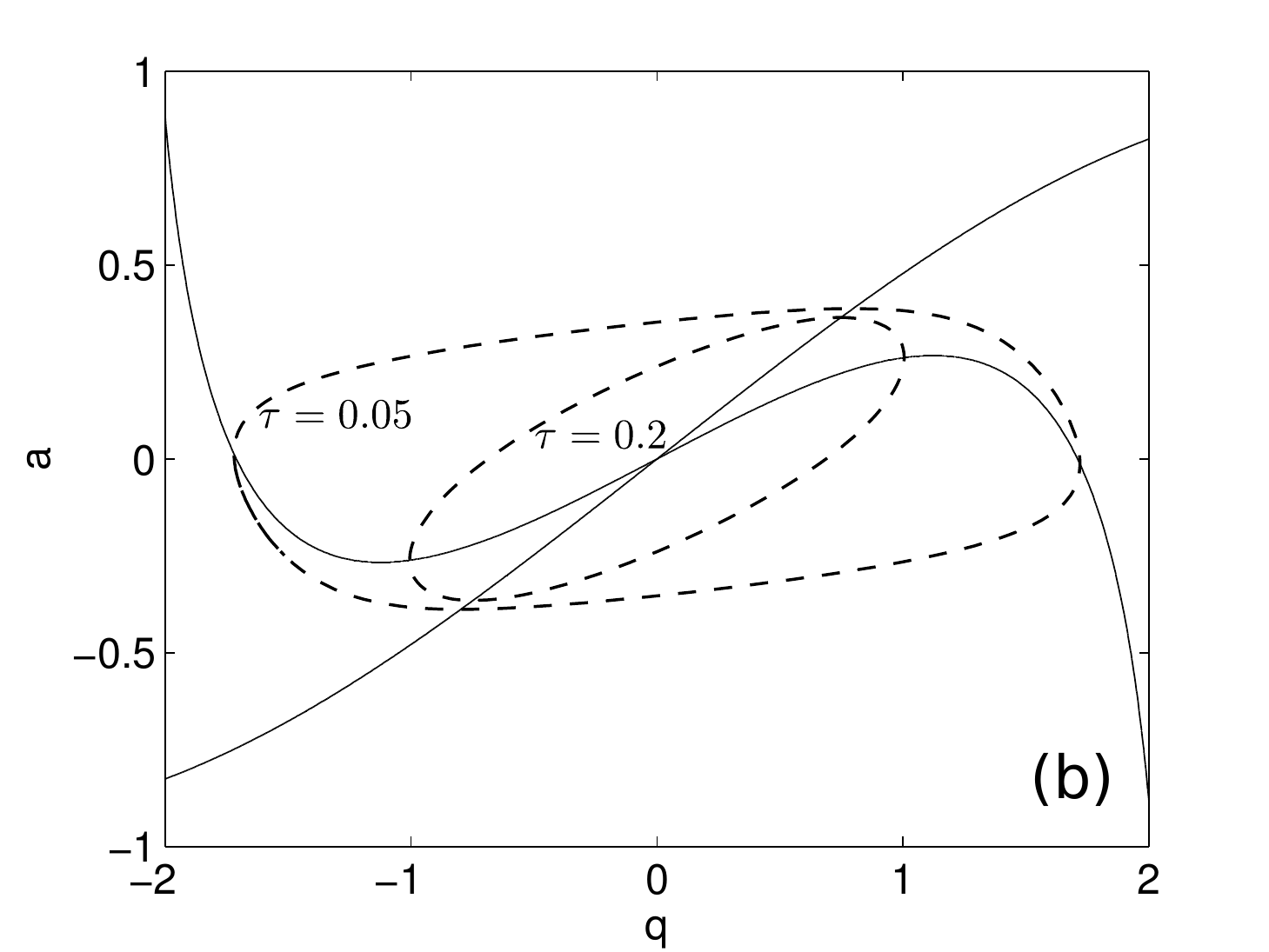}
\caption{(a) Nullclines of the system~\eqref{eq:rescaled}, $\epsilon = 0$, for $\gamma=0.5,\,R=3$. Nullcline $\dot a=0$ is reproduced twice for $\Phi=0.2$ and $\Phi=1$ corresponding respectively to regions 3 and 4 of the phase diagram in Fig.~\ref{fig:stability}. (b) Phase portraits and nullclines of the system~\eqref{eq:rescaled}, $\epsilon = 0$, for $\gamma=0.5,\,R=3,\Phi=1$. The illustrated limit cycles are obtained numerically for $\tau=0.05$ and $\tau=0.2$ during time interval of $100$ units; initial conditions were chosen randomly at the beginning of each simulation. \label{fig:nullclines}}
\end{figure*}

By analyzing the first three terms in the Maclaurin series for the right-hand side of the system~\eqref{eq:rescaled}, $\epsilon = 0$, we derived an approximation for the condition of emergence of two extra fixed points. In fact, it coincides with the inverse of the second inequality in~\eqref{eq:stability}: the system has three fixed points when $\Phi<R(1-\gamma)-1$. Thereby, inside the regions 2 and 3 the system has three equilibria: the unstable origin and two symmetrically located fixed points ($q=q^*$,\,$a=a^*$) and ($q=-q^*$,\,$a=-a^*$). Numerical simulations reveal that the latter two equilibria are both stable: the system eventually reaches one of them depending on the initial conditions. The example of such configuration for $\Phi=1$ is illustrated by the triple intersection of the system nullclines represented in Fig.~\ref{fig:nullclines}a. 

Further numerical analysis demonstrates that inside the region 4, being unstable at the origin and having no other fixed points, the system performs periodic oscillations (Fig.~\ref{fig:nullclines}b). Both the amplitude and the form of the resulting limit cycle depend on the time scale parameter $\tau$. For instance, the system exhibits relaxation oscillations for $\tau \ll 1$.
\begin{figure*}
\centering
\includegraphics[width=0.7\columnwidth]{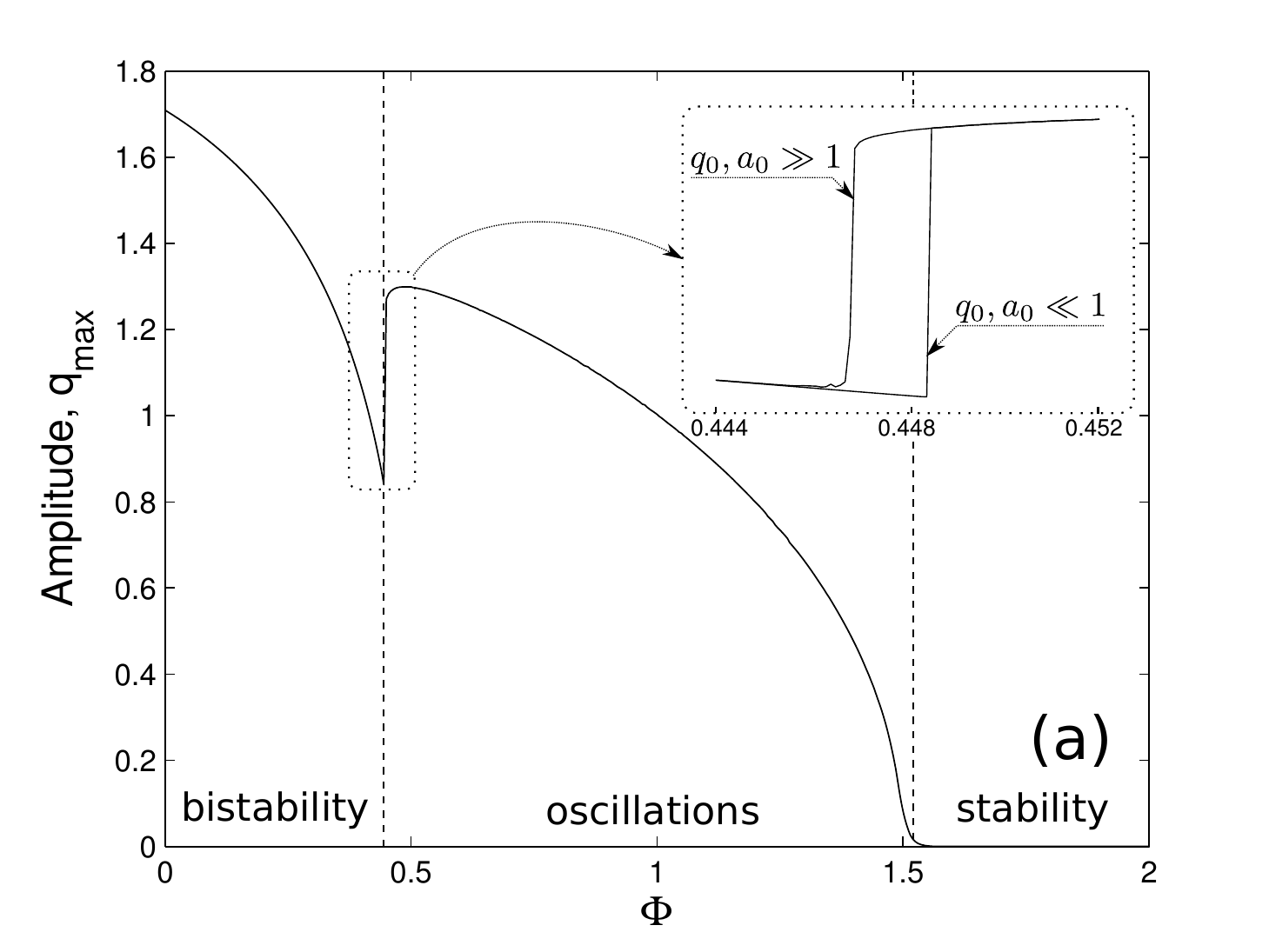}
\includegraphics[width=0.7\columnwidth]{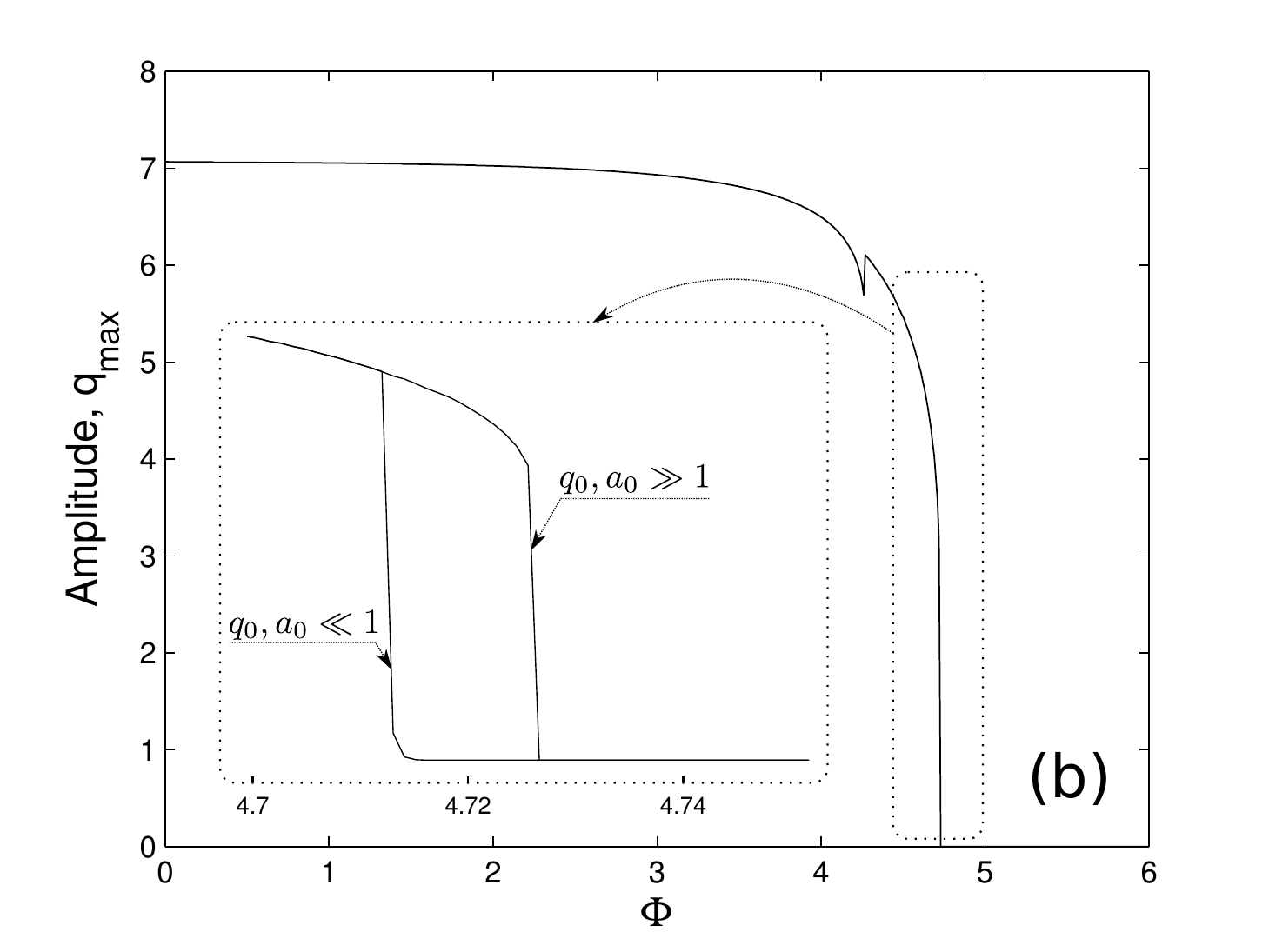}
\caption{Dependence of the maximum deviation from the origin of the system~\eqref{eq:rescaled}, $\epsilon = 0$, on parameter $\Phi$ for (a) $\tau=0.2,\,R=3,\, \gamma=0.5$; (b) $\tau=0.7,\,R=10,\, \gamma=0.5$. Maximum deviation $q_{\text{max}}$ is defined as the absolute value of the equilibrium coordinate $q$ in the case of stationary motion and the maximum value of $q$ reached by the periodic orbit in the case of oscillatory behavior. The enlarged fragments demonstrate the hystereses: the system behavior depends essentially on the initial conditions $q_0,a_0$. \label{fig:hyst}}
\end{figure*}

The deviation of the system trajectory from the equilibrium $(q=0,\,a=0)$ depending on the key parameter $\Phi$ is illustrated in Fig.~\ref{fig:hyst}. The deviation monotonically decreases with $\Phi$, except the sharp jump corresponding to the transition from bistable behavior to oscillations. The dynamics of the system~\eqref{eq:rescaled}, $\epsilon = 0$, near the boundaries of stability is complex. We discovered that for some values of $\Phi$ the system may have a hysteresis: the dynamics depend essentially on its history (Fig.~\ref{fig:hyst}). At the moment when the system ``switches'' between bistable dynamics and periodic oscillations (regions 3 and 4 in~Fig.~\ref{fig:stability}) due to variations of parameter $\Phi$, both of these patterns can be observed depending on the initial conditions (Fig.~\ref{fig:hyst}a). Furthermore, when $\tau$ approaches unity another hysteresis emerges near the border between regions 1 and 4 (Fig.~\ref{fig:hyst}b). Such behavior indicates the complexity of the system 
dynamics, still, the study of its details is left for future work.

Fig.~\ref{fig:stability} summarizes the results of the theoretical and numerical analysis of system~\eqref{eq:rescaled}, $\epsilon = 0$. The phase variable $a$ contributes to the agent preference considerably when the parameter $\Phi$ characterizing the impact of novelty seeking exceeds some threshold value. In case of absent or weak impact of novelty seeking the agent tends to permanently select the option that seemed the most attractive when the process started (one of the two stable equilibria is reached depending on the initial conditions, Fig.~\ref{fig:nullclines}a, $\Phi=0.2$). When the influence of novelty seeking becomes strong enough (that is, $\Phi>R(1-\gamma)-1$), the agent switches to one of two strategies described below depending on the value of the memory parameter $\tau$.

If $\tau$ is small in some sense, i.e., the memories within channel $\mathbb{Q}$ fade away faster comparing to that of channel $\mathbb{A}$, the agent preference begins to oscillate (Fig.~\ref{fig:nullclines}b). The mechanism behind these oscillations is rather intuitive. Right after the adaptation process starts, one of the available options inevitably becomes prevalent. If the agent estimate of the objective quality of the currently preferred alternative becomes relatively high, the corresponding choice probability approaches unity. At the same time the aversion to this option peaks, because other options, being almost never selected, appear novel. The high values of aversion cause the probability to drop, whereas the low aversion allows it to grow fast; in such manner the periodic behavior emerges. In this case the increasing impact of novelty seeking destabilizes the dynamics, causing the choice probability distribution to oscillate continuously.

If $\Phi$ is further increased, the choice probabilities distribution becomes stable. Particularly, when $\Phi$ exceeds $\frac{R(1-\gamma)-1}{\tau}-1$, the origin ($q=0,\,a=0$) turns to be stable, so the choice probability distribution is uniform: $p_1=p_2=0.5$. In other words, two options become indistinguishable and the choice is effectively random.

\subsection{The case of unequal rewards}
\begin{figure*}
\centering
\includegraphics[width=0.6\columnwidth]{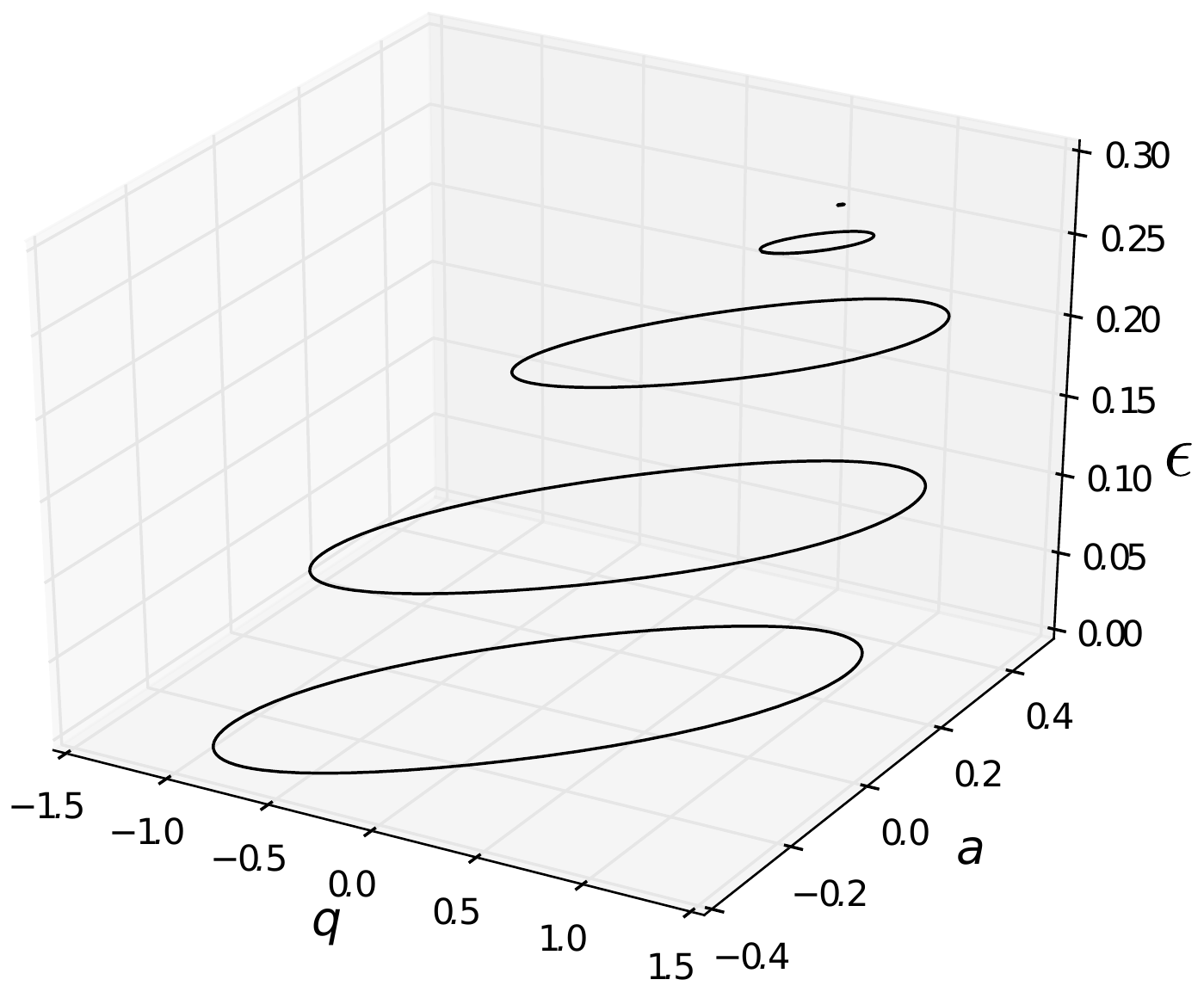}
\caption{Trajectories of the system~\eqref{eq:rescaled} for various values of the reward distortion parameter ($\epsilon \in \{ 0,\,0.1,\,0.2,\,0.25,\,0.27 \}$). The values of other parameters used for simulations are $\Phi=1,\, \tau=0.2,\,R=3,\,  \gamma=0.5$. 
\label{fig:cycles}}
\end{figure*}

Whenever the system exhibits oscillations in case of equal rewards (region 4 in Fig.~\ref{fig:stability}), the periodic behavior persists for small enough values of the reward distortion parameter $\epsilon$. The limit cycle gradually shrinks and shifts toward the higher values of $q$ and $a$ with growing $\epsilon$ (see Fig.~\ref{fig:cycles}). Finally, the limit cycle transforms to the stable equilibrium corresponding to the optimal strategy~$p_1=1$.

\begin{figure*}
\centering
\includegraphics[width=0.48\columnwidth]{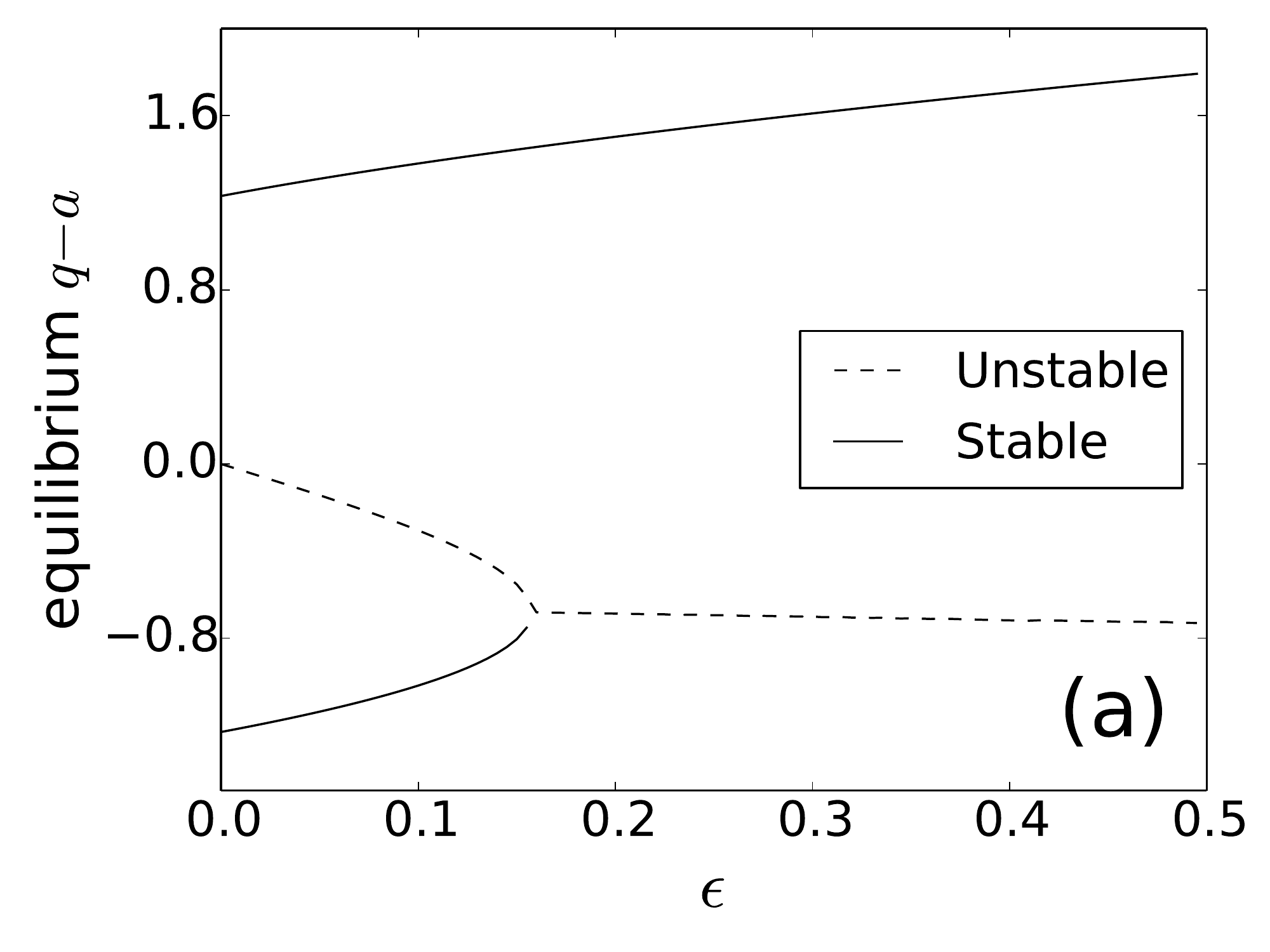}
\hspace{0.01\columnwidth}
\includegraphics[width=0.48\columnwidth]{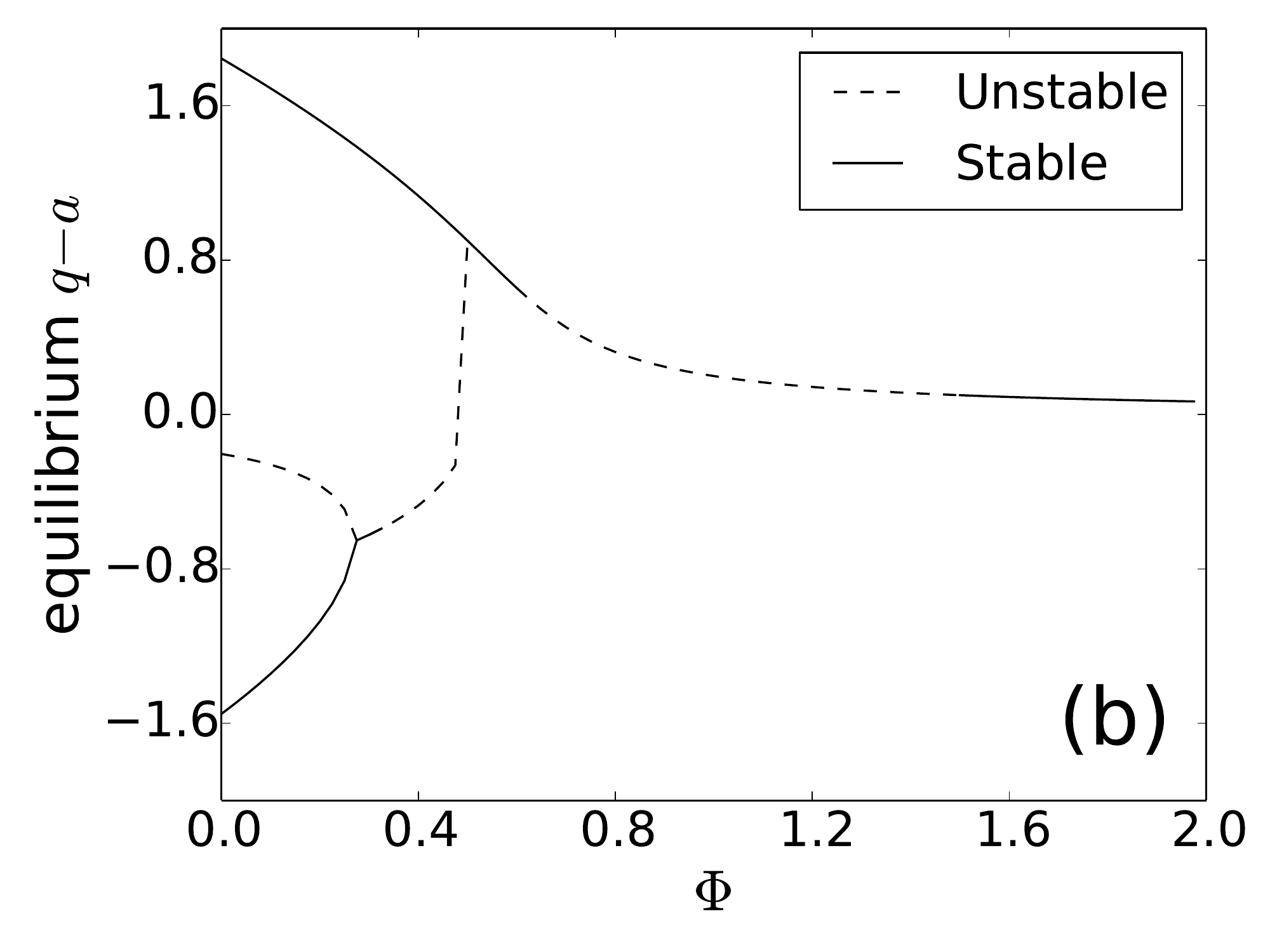} 
\caption{Bifurcation diagram of the system~\eqref{eq:rescaled} for (a) parameter $\epsilon$ (with other parameters fixed at $\Phi=0.25,\,\tau=0.2,\,R=3,\, \gamma=0.5$) and (b) parameter $\Phi$ for $\epsilon=0.1,\tau=0.2,\,R=3,\, \gamma=0.5$. The instability of the single equilibrium for $\Phi \in (0.6,\,1.5)$ corresponds to periodic oscillations around this equilibrium.
\label{fig:bifurc}}
\end{figure*}

In case that the unperturbed system is bistable (regions 2 and 3 in Fig.~\ref{fig:stability}), the non-zero difference in rewards also does not cause sudden changes in the stability of the two equilibria (see Fig.~\ref{fig:bifurc}a). This produces the paradoxical situation when for some initial values of $q$ and $a$ the agent always favors the option $x_2$, which minimizes the cumulative reward instead of maximizing it. Not surprisingly, further increase of $\epsilon$ makes this non-optimal equilibrium unstable, so that the rational strategy $p_1=1$ is adopted for practically all initial conditions. 

Fig.~\ref{fig:bifurc}b illustrates how the dynamics of the system depend on the parameter $\Phi$. For small values of $\Phi$ the adaptation process may converge to the non-optimal strategy $p_2=1$ due to specific initial conditions. Remarkably, increase of the parameter $\Phi$ leads to the choice of the optimal strategy ($q-a>0$, i.e., $p_1 \approx 1$). For some range of values of $\Phi$ (e.g., $\Phi \in (0.33,0.6)$ in the example illustrated in Fig.~\ref{fig:bifurc}b) there is a single equilibrium $q-a>0$ which is necessarily achieved regardless of the initial conditions. The magnitude of $q-a$ gradually decreases, thereby asymptotically leading to the uniformly random choice ($q-a \approx 0$). However, the transition to the stable equiprobable choice strategy is not smooth: when $\Phi$ exceeds certain threshold, the oscillatory motion emerges (the limit cycle for $\Phi=1.0,\, \epsilon=0.1$ is represented in Fig.~\ref{fig:cycles}). Further growth of $\Phi$ leads to the stability of the equilibrium~$q-
a \approx 0$.

Naturally, the presented results are highly preliminary and need thorough investigation, e.g., the scrutiny of the system hysteretic behavior and the detailed bifurcation analysis are still to be done. 

\subsection{Discussion}
\label{sec:discussion}
The analytical and numerical investigations of the system~\eqref{eq:rescaled} yield several notable facts about the system dynamics. We highlight and briefly discuss some of them appealing to the initial assumptions of the proposed model.
\begin{enumerate}
\item The model can capture the non-trivial dynamics (namely, unstable oscillations) of the single-agent adaptation in case of two equally rewarded actions. This situation may be linked to the experimental findings on experience-based decision making, where the preference of human subjects fluctuate considerably around the Nash equilibrium in the repeated choice task with two alternatives~\cite{barron2003small}. 
\item When the symmetry of the two actions is distorted, the oscillations persist, so the agent behavior may be unstable even when the optimal strategy exists.
\item In case of unequal rewards varying the extent to which novelty seeking affects the agent choice may lead to surprising changes in the system dynamics:
\begin{enumerate}
\item When novelty seeking has little effect on the agent choice, the adaptation may stagnate at the non-optimal option. Still, this may happen only for rather small difference between the two options. The minor difference in rewards is not enough to motivate the agent initially inclined to the inferior alternative to switch to the optimal one. In other words, the lack of novelty seeking may hinder the effective adaptation process.
\item Growing impact of novelty seeking makes the agent able to overcome the initial predisposition and successfully learn the optimal strategy. It can be interpreted in a way that the increasing influence of the novelty seeking trait may improve the performance of the adaptation and thereby make the agent more rational in the end.
\item However, further increase of the relative impact of novelty seeking makes the agent choice inconsistent. The interplay between the reward-seeking and novelty-seeking motives may cause the choice probabilities to periodically oscillate. Novelty seeking thus appears as a source of emergence of adaptation instability.
\item When the impact of novelty seeking becomes in some sense higher than that of reward seeking, the rewards are effectively neglected by the agent. All the options become practically equivalent, so the agent who seeks mainly for novelty ultimately comes to the uniformly random choice.
\end{enumerate}
\end{enumerate}

The de facto standard single-agent learning model can reproduce only the most basic, rational behavior pattern: given enough memory capacity, the agent generally learns the optimal behavior strategy and then strictly follows it~\cite{sato2005stability}. In terms of dynamics it means that for all plausible values of system parameters the optimal strategy is a stable equilibrium point in the space of all possible strategies. According to the preliminary analysis, the model presented here can reproduce much more diverse patterns of the agent behavior. The captured dynamical behaviors seem psychologically plausible at first glance, although the model has not yet been directly confronted with the experimental studies on human adaptation. 

Indeed, many factors other than novelty seeking may hypothetically impact on the adaptation dynamics. For instance, the proposed extension of the reinforcement learning framework can be used to capture the stability-seeking behavior, when the frequently selected options are preferred by the agent. It is likely that in many applications both stability seeking and novelty seeking can be pronounced and may even coexist. Nevertheless, from the dynamical perspective the effect of the stability-seeking behavior is of much less interest. The bias towards the frequently selected actions will most likely lead to the stagnation of the adaptation at the initially preferred alternative. On the other hand, the factors that drive the agent towards the rarely chosen, novel actions seem more feasible as an emergence mechanism leading to rich adaptation dynamics. 

\section{Conclusion}
We propose a dynamical model of adaptation in unknown environment under effect of novelty seeking. In the conventional models used in the non-equilibrium game theory the learning agents are assumed to act rationally in achieving the ultimate goal \textemdash to maximize the cumulative reward gained during the learning. Such models lead to complex phenomena only in case of multiple interacting agents, while the single agent adaptation has been generally presumed to be trivial up to now.

We challenge this approach by endowing the learning agent with two interacting processing channels. The standard reward-based learning is handled by the channel $\mathbb{Q}$ whereas the channel $\mathbb{A}$ is associated with the agent's attraction to novel actions. We confine our scope to the case of the single agent adaptation and demonstrate that novelty seeking may lead to the instability of the learning dynamics: the intermediate levels of novelty seeking cause the choice probability distribution to oscillate. Moreover, we discover that the agent characterized by the moderate impact of novelty seeking can be in a certain sense more rational comparing to the agent whose choice is governed strictly by rewards. 

Our results demonstrate that accounting for more mental complexity of the agent greatly extends the spectrum of dynamical behaviors captured by the reinforcement learning model and can enable the basic explanation of transitions between these behaviors depending on the parameters.

The possible flaw of our highly theoretical analysis is that the psychological plausibility of the parameter values under which the instability is achieved (i.e., $\tau < 1$ and $0<\gamma<1$) remains an open question. Nonetheless, we believe that the presented model may serve as an overall psychologically adequate alternative to the standard reinforcement learning model, at the same time accounting for essentially more diverse dynamical phenomena.

The presented model allows for a number of extensions. For instance, it may aid in studying the effects of the diverse heuristics that humans employ in judgements and decision making~\cite{tversky1974judgment} on the dynamics of learning. Another open question is how exactly would all the discussed effects change the dynamics of the interaction between multiple agents comprising a complex system? It is entirely possible that novelty seeking (and other effects of the similar nature) would not in general prove itself as an emergence mechanism and would not enrich the collective adaptation dynamics. Still, we believe that this question is worth exploring. It is generally agreed that personality traits greatly impact on human behavior via a variety of factors~\cite{cloninger2004feeling}. Establishing a connection between such factors and emergent social phenomena is an open problem for both psychology and computational social science.

\section*{Acknowledgments}
The work was supported in part by the JSPS ``Grants-in-Aid for Scientific Research'' Program, Grant 24540410-0001.

\bibliographystyle{ws-acs}
\bibliography{library}

\begin{thebibliography}{10}
\providecommand{\urlprefix}{}
\expandafter\ifx\csname urlstyle\endcsname\relax
  \providecommand{\doi}[1]{doi:\discretionary{}{}{}#1}\else
  \providecommand{\doi}{doi:\discretionary{}{}{}\begingroup
  \urlstyle{rm}\Url}\fi

\bibitem{arthur1994inductive}
Arthur, W.~B., {Inductive reasoning and bounded rationality}, \emph{The
  American Economic Review} \textbf{84} (1994) 406--411.

\bibitem{barron2003small}
Barron, G. and Erev, I., {Small feedback-based decisions and their limited
  correspondence to description-based decisions}, \emph{Journal of Behavioral
  Decision Making} \textbf{16} (2003) 215--233.

\bibitem{borgers1997learning}
B{\"o}rgers, T. and Sarin, R., {Learning through reinforcement and replicator
  dynamics}, \emph{Journal of Economic Theory} \textbf{77} (1997) 1--14.

\bibitem{cloninger1985unified}
Cloninger, C.~R., {A unified biosocial theory of personality and its role in
  the development of anxiety states.}, \emph{Psychiatric developments}
  \textbf{4} (1985) 167--226.

\bibitem{cloninger2004feeling}
Cloninger, C.~R., \emph{{Feeling good: the science of well-being}} (Oxford
  University Press, 2004).

\bibitem{cloninger1993psychobiological}
Cloninger, C.~R., Svrakic, D.~M., and Przybeck, T.~R., {A psychobiological
  model of temperament and character}, \emph{Archives of general psychiatry}
  \textbf{50} (1993) 975.

\bibitem{conte2012manifesto}
Conte, R., Gilbert, N., Bonelli, G., Cioffi-Revilla, C., Deffuant, G., Kertesz,
  J., Loreto, V., Moat, S., Nadal, J.-P., Sanchez, A., \emph{et~al.},
  {Manifesto of computational social science}, \emph{The European Physical
  Journal Special Topics} \textbf{214} (2012) 325--346.

\bibitem{deci1985intrinsic}
Deci, E. and Ryan, R., \emph{{Intrinsic motivation and self-determination in
  human behavior}} (Springer, 1985).

\bibitem{deneubourg1999self}
Deneubourg, J.-L., Camazine, S., and Detrain, C., {Self-organization or
  individual complexity: a false dilemma or a true complementarity?}, in
  \emph{{Information processing in social insects}} (Springer, 1999), pp.
  401--407.

\bibitem{erev1998predicting}
Erev, I. and Roth, A.~E., {Predicting how people play games: Reinforcement
  learning in experimental games with unique, mixed strategy equilibria},
  \emph{American Economic Review}  (1998) 848--881.

\bibitem{fudenberg1998theory}
Fudenberg, D. and Levine, D., \emph{{The theory of learning in games}} (MIT
  press, 1998).

\bibitem{galla2009intrinsic}
Galla, T., {Intrinsic noise in game dynamical learning}, \emph{Physical Review
  Letters} \textbf{103} (2009) 198702.

\bibitem{galla2011cycles}
Galla, T., {Cycles of cooperation and defection in imperfect learning},
  \emph{Journal of Statistical Mechanics: Theory and Experiment} \textbf{2011}
  (2011) P08007.

\bibitem{ho2008individual}
Ho, T.~H., Wang, X., and Camerer, C.~F., {Individual differences in EWA
  learning with partial payoff information}, \emph{The Economic Journal}
  \textbf{118} (2008) 37--59.

\bibitem{kianercy2012dynamics}
Kianercy, A. and Galstyan, A., {Dynamics of Boltzmann Q learning in two-player
  two-action games}, \emph{Physical Review E} \textbf{85} (2012) 041145.

\bibitem{kuhnen2009genetic}
Kuhnen, C.~M. and Chiao, J.~Y., {Genetic determinants of financial risk
  taking}, \emph{PLoS One} \textbf{4} (2009) e4362.

\bibitem{lee1992measuring}
Lee, T.-H. and Crompton, J., {Measuring novelty seeking in tourism},
  \emph{Annals of tourism research} \textbf{19} (1992) 732--751.

\bibitem{leslie2005individual}
Leslie, D.~S. and Collins, E., {Individual Q-learning in normal form games},
  \emph{SIAM Journal on Control and Optimization} \textbf{44} (2005) 495--514.

\bibitem{lubashevsky2010scale}
Lubashevsky, I. and Kanemoto, S., {Scale-free memory model for multiagent
  reinforcement learning. Mean field approximation and rock-paper-scissors
  dynamics}, \emph{The European Physical Journal B-Condensed Matter and Complex
  Systems} \textbf{76} (2010) 69--85.

\bibitem{macy2002learning}
Macy, M. and Flache, A., {Learning dynamics in social dilemmas},
  \emph{Proceedings of the National Academy of Sciences} \textbf{99} (2002)
  7229--7236.

\bibitem{oudeyer2007what}
Oudeyer, P. and Kaplan, F., {What is intrinsic motivation? a typology of
  computational approaches}, \emph{Frontiers in Neurorobotics} \textbf{1}
  (2007).

\bibitem{oudeyer2007intrinsic}
Oudeyer, P., Kaplan, F., and Hafner, V., {Intrinsic motivation systems for
  autonomous mental development}, \emph{Evolutionary Computation, IEEE
  Transactions on} \textbf{11} (2007) 265--286.

\bibitem{paulus2012emotion}
Paulus, M.~P. and Yu, A.~J., {Emotion and decision-making: affect-driven belief
  systems in anxiety and depression}, \emph{Trends in cognitive sciences}
  (2012).

\bibitem{roces2002individual}
Roces, F., {Individual complexity and self-organization in foraging by
  leaf-cutting ants}, \emph{The Biological Bulletin} \textbf{202} (2002)
  306--313.

\bibitem{rushworth2008choice}
Rushworth, M.~F. and Behrens, T.~E., {Choice, uncertainty and value in
  prefrontal and cingulate cortex}, \emph{Nature neuroscience} \textbf{11}
  (2008) 389--397.

\bibitem{ryan2000intrinsic}
Ryan, R. and Deci, E., {Intrinsic and extrinsic motivations: Classic
  definitions and new directions}, \emph{Contemporary educational psychology}
  \textbf{25} (2000) 54--67.

\bibitem{ryan2000self}
Ryan, R.~M. and Deci, E.~L., {Self-determination theory and the facilitation of
  intrinsic motivation, social development, and well-being}, \emph{American
  psychologist} \textbf{55} (2000) 68--78.

\bibitem{sato2005stability}
Sato, Y., Akiyama, E., and Crutchfield, J., {Stability and diversity in
  collective adaptation}, \emph{Physica D: Nonlinear Phenomena} \textbf{210}
  (2005) 21--57.

\bibitem{sato2002chaos}
Sato, Y., Akiyama, E., and Farmer, J., {Chaos in learning a simple two-person
  game}, \emph{Proceedings of the National Academy of Sciences} \textbf{99}
  (2002) 4748--4751.

\bibitem{sato2003coupled}
Sato, Y. and Crutchfield, J., {Coupled replicator equations for the dynamics of
  learning in multiagent systems}, \emph{Physical Review E} \textbf{67} (2003)
  015206.

\bibitem{tversky1974judgment}
Tversky, A. and Kahneman, D., {Judgment under Uncertainty: Heuristics and
  Biases}, \emph{Science} \textbf{185} (1974) 1124--1131.

\end{thebibliography}

\end{document}